\documentclass[10pt,letterpaper]{article}
\usepackage{opex3}

\usepackage{amsmath,amsfonts,amssymb}
\usepackage{graphicx}
\usepackage{enumerate}
\usepackage{bbm}
\usepackage{color}
\usepackage{epstopdf}
\usepackage{calc}

\newcommand{\eg}{\emph{e.g.}}
\newcommand{\ie}{\emph{i.e.}}

\newcommand{\ii}{\ensuremath{\text{i}}}

\DeclareMathOperator{\erf}{erf}

\begin{document}
 
\title{Conformal carpet and grating cloaks}

\author{Roman Schmied,$^{1,*}$ Jad C.~Halimeh,$^{2,3}$ and Martin Wegener$^4$}
\address{$^1$Departement Physik, Universit\"at Basel, Switzerland\\
$^2$Max-Planck-Institut f\"ur Quantenoptik, Garching, Germany\\
$^3$Fakult\"at f\"ur Physik, Ludwig-Maximilians-Universit\"at, M\"unchen, Germany\\
$^4$Institut f\"ur Angewandte Physik, DFG Center for Functional Nanostructures (CFN), and Institut f\"ur Nanotechnologie, Karlsruhe Institute of Technology, Germany}
\email{$^*$Roman.Schmied@unibas.ch}
 
\begin{abstract}
We introduce a class of conformal versions of the previously introduced quasi-conformal carpet cloak, and show how to construct such conformal cloaks for different cloak shapes. Our method provides exact refractive-index profiles in closed mathematical form for the usual carpet cloak as well as for other shapes. By analyzing their asymptotic behavior, we find that the performance of finite-size cloaks becomes much better for metal shapes with zero average value, \eg, for gratings.
\end{abstract}
 

\bibliographystyle{osajnl}
\bibliography{MPQ}

\begin{thebibliography}{10}
\newcommand{\enquote}[1]{``#1''}

\bibitem{Li2008}
J.~Li and J.~B. Pendry, \enquote{Hiding under the carpet: A new strategy for
  cloaking,} Phys. Rev. Lett. \textbf{101}, 203901 (2008).

\bibitem{Liu2009}
R.~Liu, C.~Ji, J.~J. Mock, J.~Y. Chin, T.~J. Cui, and D.~R. Smith,
  \enquote{Broadband ground-plane cloak,} Science \textbf{323}, 366--369
  (2009).

\bibitem{Valentine2009}
J.~Valentine, J.~Li, T.~Zentgraf, G.~Bartal, and X.~Zhang, \enquote{An optical
  cloak made of dielectrics,} Nature Mater. \textbf{8}, 568--571 (2009).

\bibitem{Gabrielli2009}
L.~H. Gabrielli, J.~Cardenas, C.~B. Poitras, and M.~Lipson, \enquote{Silicon
  nanostructure cloak operating at optical frequencies,} Nature Photon.
  \textbf{3}, 461--463 (2009).

\bibitem{Lee2009}
J.~H. Lee, J.~Blair, V.~A. Tamma, Q.~Wu, S.~J. Rhee, C.~J. Summers, and
  W.~Park, \enquote{Direct visualization of optical frequency invisibility
  cloak based on silicon nanorod array,} Opt. Express \textbf{17}, 12922--12928
  (2009).

\bibitem{Ergin2010}
T.~Ergin, N.~Stenger, P.~Brenner, J.~B. Pendry, and M.~Wegener,
  \enquote{Three-dimensional invisibility cloak at optical wavelengths,}
  Science \textbf{328}, 337--339 (2010).

\bibitem{Ma2010}
H.~F. Ma and T.~J. Cui, \enquote{Three-dimensional broadband ground-plane cloak
  made of metamaterials,} Nature Commun. \textbf{1}, 1--6 (2010).

\bibitem{Zhang2010}
B.~Zhang, T.~Chan, and B.-I. Wu, \enquote{Lateral shift makes a ground-plane
  cloak detectable,} Phys. Rev. Lett. \textbf{104}, 233903 (2010).

\bibitem{Nicorovici2008}
N.-A.~P. Nicorovici, R.~C. McPhedran, S.~Enoch, and G.~Tayeb, \enquote{Finite
  wavelength cloaking by plasmonic resonance,} New J. Phys. \textbf{10}, 115020
  (2008).

\bibitem{Shalaev2008}
V.~M. Shalaev, \enquote{Transforming light,} Science \textbf{322}, 384--386
  (2008).

\bibitem{Chen2010}
H.~Chen, C.~T. Chan, and P.~Sheng, \enquote{transformation optics and
  metamaterials,} Nature Mater. \textbf{9}, 387--396 (2010).

\bibitem{Zhang2010b}
P.~Zhang, M.~Lobet, and S.~He, \enquote{Carpet cloaking on a dielectric
  half-space,} Opt. Express \textbf{18}, 15158--18163 (2010).

\bibitem{Leonhardt2006}
U.~Leonhardt, \enquote{Optical conformal mapping,} Science \textbf{312},
  1777--1780 (2006).

\end{thebibliography}

\section{Introduction}

Soon after the introduction of the so-called carpet cloak by Li and Pendry in 2008~\cite{Li2008}, broadband invisibility cloaking has become experimental reality in two~\cite{Liu2009,Valentine2009,Gabrielli2009,Lee2009} and three~\cite{Ergin2010,Ma2010} dimensions from microwaves~\cite{Liu2009,Ma2010} to the optical regime~\cite{Valentine2009,Gabrielli2009,Lee2009,Ergin2010}. In essence, the carpet or ground-plane cloak makes a bump (more generally a corrugation) in a metallic carpet appear flat and hence undetectable. Objects may be hidden in the space underneath the bump. However, Ref.~\cite{Zhang2010} has highlighted lateral beam displacements as an inherent limitation originating from approximating the locally anisotropic optical properties, which arise from the quasi-conformal mappings employed in the construction of these finite-size cloaks, by locally \emph{isotropic} ones. This artifact contributes to the Ostrich effect~\cite{Nicorovici2008}: one cannot see the cloaked bump, but one can see that \emph{something} is there. These limitations can be traced back to the finite size of the cloak, since for an infinite carpet cloak the quasi-conformal map becomes strictly conformal, in which case cloaking becomes perfect in both wave and ray optics (for reviews on transformation optics see Refs.~\cite{Shalaev2008,Chen2010}).

Recently, a specific example for a strictly conformal map has been given and discussed~\cite{Zhang2010b}. Here we introduce an entire class of strictly conformal maps and discuss how the performance of \emph{finite-size} carpet cloaks can be systematically improved with respect to Refs.~\cite{Zhang2010} and~\cite{Zhang2010b}.

\section{Conformal mapping}

Let us start by emphasizing that our approach follows a somewhat different spirit than the one by Li and Pendry~\cite{Li2008}. They start with a predefined shape of the bump. In principle, its shape as well as its aspect ratio can be chosen arbitrarily. Furthermore, they fix the boundaries of the finite-size cloaking structure (with slipping boundary conditions). Numerically minimizing the modified-Liao functional~\cite{Li2008}, they arrive at a quasi-conformal map -- the closest one can get to a conformal map under the given constraints. We rather introduce a class of \emph{strictly conformal} transformations. For each transformation of this class, the shape of the bump results automatically, \ie, it can generally \emph{not} be chosen arbitrarily. However, in the special and rather important limit of shallow bumps (which \emph{all} of the aforementioned experiments have used~\cite{Liu2009,Valentine2009,Gabrielli2009,Lee2009,Ergin2010,Ma2010}), the connection between bump shape and conformal transformation is mathematically simple and intuitive. In this case, the bump shape can again be chosen arbitrarily. Furthermore, in our approach the ideal cloaking structure is infinitely extended; it has no intrinsic boundaries. One can, however, just truncate the refractive-index profile to obtain a finite-size cloak. In this paper we show that under certain constraints the effect of this truncation can be minimized.

Mathematically, we start from the conformal map $z \mapsto f(z)$ given by
\begin{equation}
	\label{eq:conformalmap}
	f(z) = z + \int_0^{\infty} c_k e^{\ii k z}\text{d}k,
\end{equation}
where $z=x+\ii y$; $x\in\mathbbm{R}$ and $y\in\mathbbm{R}^+$ are the coordinates of points in the Cartesian two-dimensional half-space above the horizontal axis, assuming translational invariance in the third dimension. They are mapped onto the transformed coordinates $(u(x,y),v(x,y))\in\mathbbm{R}^2$ given by $f(x+\ii y)=u+\ii v$. The right-hand side of Eq.~\eqref{eq:conformalmap} contains a truncated (lower integral bound is zero) Fourier transform of the coefficients $c_k$. This truncation reflects the fact that we consider only light propagating in the half-space above the metallic bump. The refractive indices of the \emph{virtual} space $n_0(x,y)$ ($=1$ in our work) and of the \emph{physical} space $n(u,v)$ are related by~\cite{Leonhardt2006}
\begin{equation}
	\label{eq:refindex}
	n(f(z))= \frac{n_0(z)}{| \text{d}f/\text{d}z |}.
\end{equation}
This form has a known \cite{Leonhardt2006} and intuitive physical interpretation: The conformal map $z\mapsto f(z)$ locally stretches (or compresses) space while preserving angles and the shapes of infinitesimally small figures. The factor by which space is stretched by the map is given by the modulus of the spatial derivative of the map, $\mathcal{C}=|f'(z)|$. But if physical space is stretched by a linear scale factor $\mathcal{C}$ with respect to virtual space, the refractive index in physical space has to be multiplied by $1/\mathcal{C}$ such that the optical path lengths are identical in the virtual and physical spaces (by Fermat's principle).

\newcommand{\legwidth}{0.3\textwidth}
\begin{figure}[h]
	\begin{center}
		\rule{0.5\textwidth}{0mm}\includegraphics[width=\legwidth]{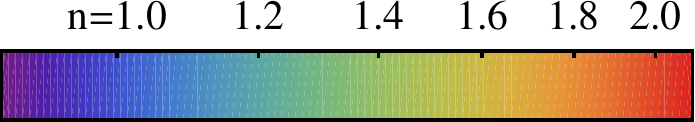}\\
		\includegraphics[width=0.49\textwidth]{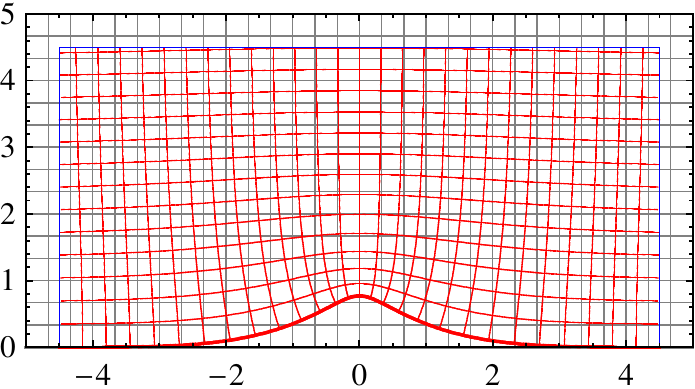}
		\includegraphics[width=0.49\textwidth]{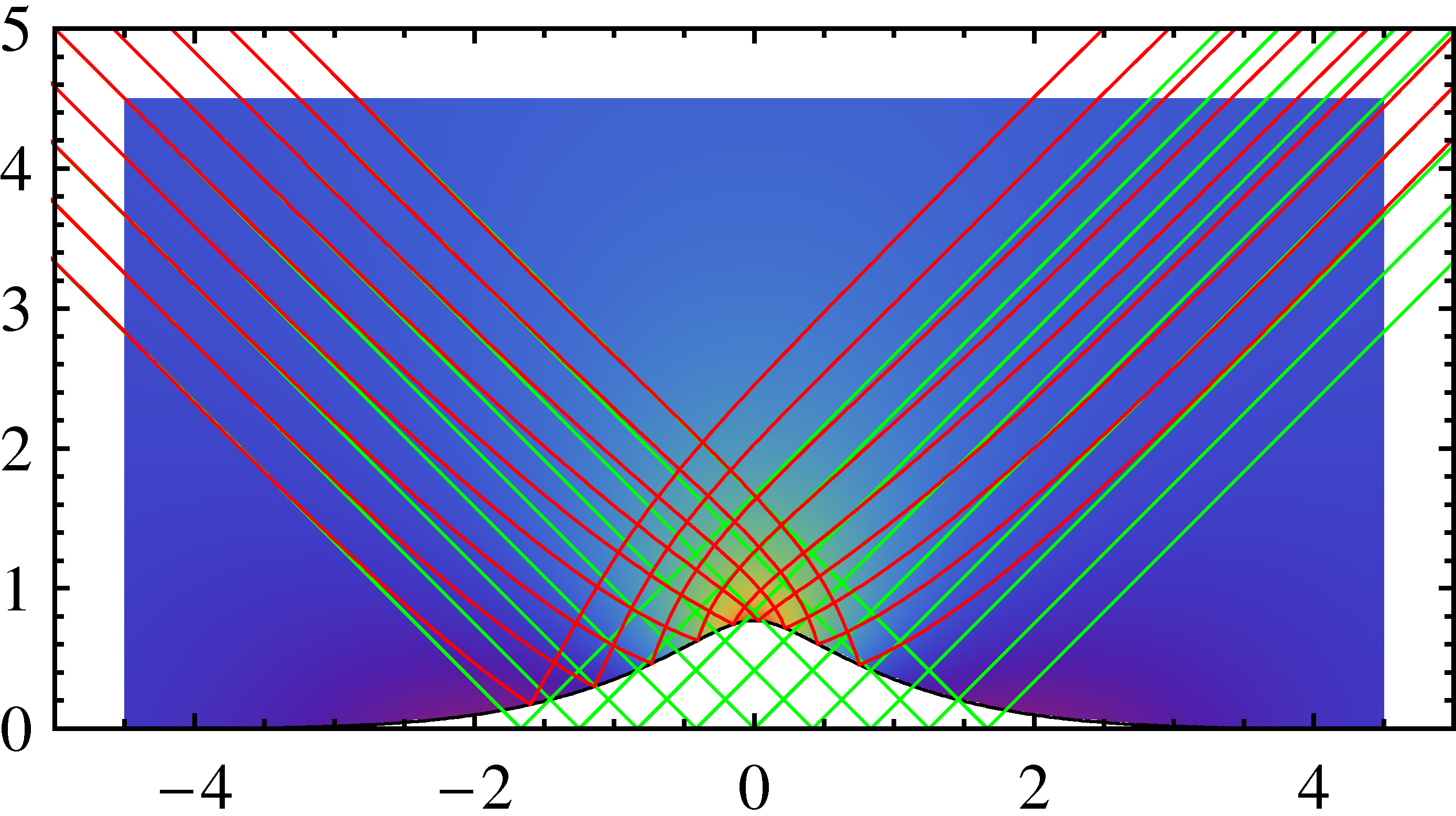}
	\end{center}
	\caption{Left: illustration of the original coordinates $(x,y)$ (gray) and the transformed coordinates $(u,v)$ (red) resulting from the cloak of Eq.~\eqref{eq:GaussianBumpCM} for $w=1.66$ and $h=0.77$ (matching the width and height of the cloak in Ref.~\cite{Zhang2010b}, see Fig.~\ref{fig:ZhangBump}). All coordinates are in normalized units. Right: refractive-index profile and selected rays. The green rays correspond to vacuum and the  mirror plane at $y=0$, the red ones to the finite cloaking structure and the ground-plane shape shown in black. The cloak has a size of $9\times4.5$ normalized units. Outside of it, we assume vacuum (white). The color scale is logarithmic, ranging from 0.86 to 2.10.}
	\label{fig:GaussBump}
\end{figure}

The shape of the bump giving rise to the refractive-index profile following from Eq.~\eqref{eq:refindex} is defined by the implicit form $(u(x,0),v(x,0))$. In general, it is difficult to obtain a closed expression for the parametric dependence of $v(x,0)$ on $u(x,0)$. However, for shallow bumps an explicit expression can be obtained. To see this, let us consider the example of a Gaussian for the coefficients $c_k$ in Eq.~\eqref{eq:conformalmap},
\begin{equation}
	\label{eq:GaussianCoeff}
	c_k = \frac{\ii h w}{\sqrt{\pi}} e^{-(k w/2)^2}\,.
\end{equation}
The conformal map is then
\begin{equation}
	\label{eq:GaussianBumpCM}
	f(z) = z + \ii h e^{-(z/w)^2} \left[ 1+\erf(\ii z/w) \right]\,,
\end{equation}
where $\erf$ is the error function (see the left panel of Fig.~\ref{fig:GaussBump}). For shallow bumps ($h \ll w$ in this example) we obtain $u(x,0) \approx x$, and derive the explicit form for the bump shape
\begin{equation}
	v(u)\approx h e^{-(u/w)^2}\,.
\end{equation}
The parameter $h$ is therefore the height of the Gaussian bump, and $w$ is its width. In general, in this limit of shallow bumps, the 
$c_k=a_k+ \ii b_k$ are the Fourier coefficients of the bump shape since
\begin{equation}
	v(x,0) = \int_0^{\infty} \left[ a_k \sin(k x) + b_k \cos(k x) \right]\text{d}k
\end{equation}
and hence, since $u(x,0)\approx x$,
\begin{equation}
	\label{eq:vofu}
	v(u) \approx \int_0^{\infty} \left[ a_k \sin(k u) + b_k \cos(k u) \right]\text{d}k\,.
\end{equation}
Thus, in this limit the coefficients $c_k$ for any desired conformal map (and refractive-index profile) can be obtained by Fourier transformation of the real-space bump shape $v(u)$.

Outside of the shallow-bump approximation, the same coefficients can still be used for perfect cloaking, but these coefficients and the corresponding conformal map $z\mapsto f(z)$ refer to a different bump shape $v(u)$ than in the shallow-bump limit. In the above example~\eqref{eq:GaussianBumpCM}, at the critical ratio $h/w=\sqrt{\pi}/2$ the maximum of the bump in the metal carpet develops into a sharp tip and the refractive index becomes singular.

We have extensively tested the above analytical results by numerical ray-tracing calculations in two dimensions. For the artificial and experimentally irrelevant case of an infinitely extended structure, cloaking is perfect -- as can be expected from the fact that the underlying transformation is strictly conformal. Experimentally relevant finite-size refractive-index profiles can be obtained by simply truncating the exact refractive-index profile, \ie, by setting $n=1$ outside of the finite-size cloak. This procedure delivers cloaking results which are similar to those obtained for the quasi-conformal carpet cloak~\cite{Zhang2010}. In particular, as can be seen on the right-hand side of Fig.~\ref{fig:GaussBump}, we obtain the same lateral beam shifts as discussed in Ref.~\cite{Zhang2010}. While the performances of quasi-conformal and conformal cloaks are qualitatively and quantitatively similar, we emphasize that our refractive-index profiles are given in closed mathematical form, whereas those of quasi-conformal carpet cloaks are derived from a nontrivial numerical minimization of the modified-Liao functional~\cite{Li2008}.

To further investigate the cloaking imperfections arising from the spatial truncation, it is interesting to study the asymptotic decay of the refractive-index profile towards its vacuum value $n=1$ far away from the bump. At large distances the refractive index due to Eq.~\eqref{eq:GaussianBumpCM} behaves like
\begin{equation}
	n(u+\ii v=\rho e^{\ii\varphi}) = 1 - \frac{h w \cos(2\varphi)}{\sqrt{\pi}\rho^2} + \mathcal{O}(\rho^{-4})\,.
\end{equation}
This decay is polynomial, which means that $n(u,v)$ approaches the vacuum limit $n=1$ rather slowly -- necessitating undesirably large cloaking structures. This slow decay, which corresponds to small spatial-frequency components in $f(z)$, is connected to the small spatial-frequency components of the bump shape $v(u)$ itself. 

\begin{figure}[h]
	\begin{center}
		\rule{0.5\textwidth}{0mm}\includegraphics[width=\legwidth]{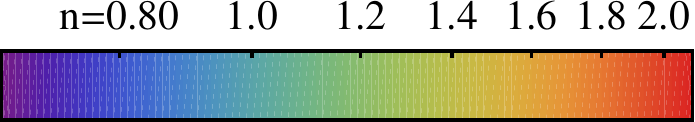}\\
		\vspace{1.1mm}
		\includegraphics[width=0.49\textwidth]{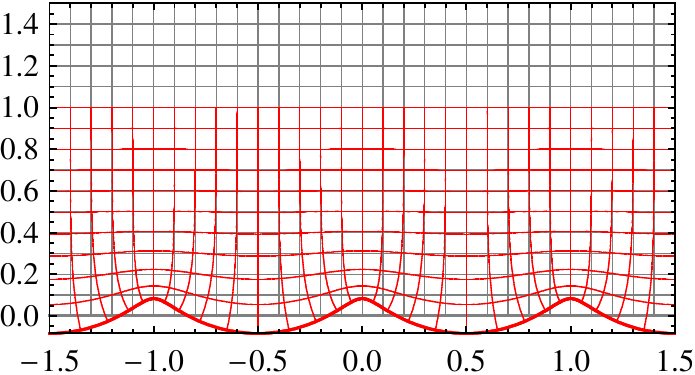}
		\includegraphics[width=0.49\textwidth]{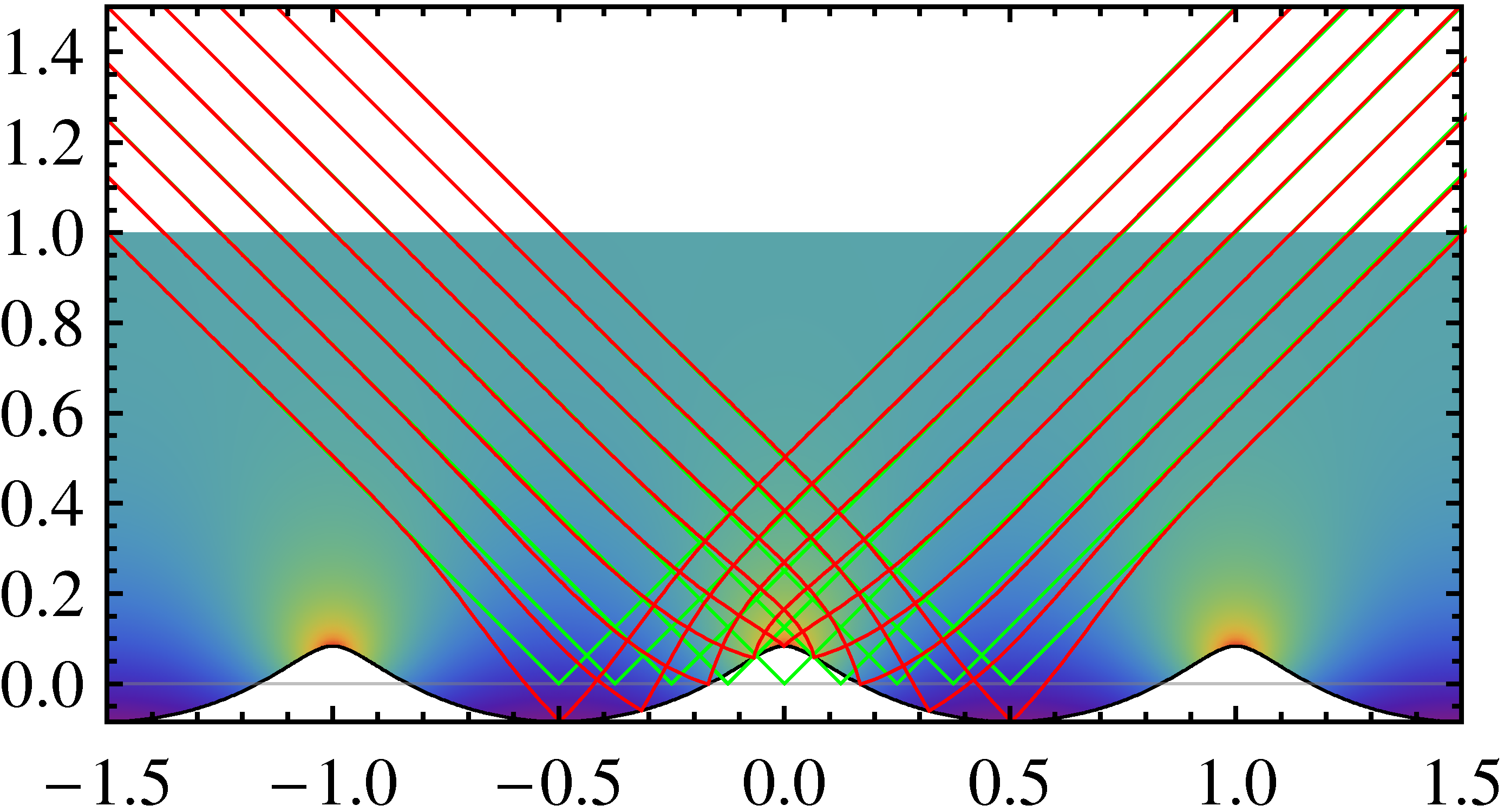}
	\end{center}
	\caption{Left: illustration of the original coordinates $(x,y)$ (gray) and the transformed coordinates $(u,v)$ (red) resulting from Eq.~\eqref{eq:wiggleCM} with $c=\ii/12$ and $k=2\pi$. All coordinates are in normalized units. Right: refractive-index profile and selected rays. The green rays correspond to vacuum and the  mirror plane indicated in gray, the red ones to the finite cloaking structure and the ground-plane shape shown in black. The cloak has a height of one normalized unit. Above it, we assume vacuum (white). The color scale is logarithmic, ranging from 0.66 to 2.10.}
	\label{fig:GaussianBumpwiggle}
\end{figure}

\newpage
To exemplify this observation, we consider a mathematically much simpler case which uses only a single non-zero spatial-frequency component $k$ in Eq.~\eqref{eq:conformalmap},
\begin{equation}
	\label{eq:wiggleCM}
	f(z)=z+c e^{\ii k z}
\end{equation}
with $c=a+\ii b$.
The resulting transformation is illustrated in the left panel of Fig.~\ref{fig:GaussianBumpwiggle}. Its refractive-index profile is shown in the right panel,
\begin{equation}
	n(\zeta=u+\ii v) =
	\left| 1 + W_0\left(\ii c k e^{\ii k \zeta} \right) \right|^{-1}
	= 1+[a\sin(ku)+b\cos(ku)]k e^{-kv}+\mathcal{O}(e^{-2kv}),
\end{equation}
where $W_0$ is the Lambert function.\footnote{The Lambert function (or product logarithm) $W_0(z)$ is the principal solution for $w$ in $z=w e^w$.}
The refractive index approaches the vacuum limit $n=1$ \emph{exponentially} fast with increasing vertical coordinate $v$. The absence of zero and small spatial-frequency components means that the average value of the ``carpet'' shape $v(u)$ is zero. This implies that the shape $v(u)$ no longer only exhibits values \emph{above} the fictitious ground plane (maxima), but also values \emph{below} that ground plane (minima) -- in sharp contrast to the usual carpet \cite{Li2008} with a single maximum. We have rather found a cloak for a corrugated metal surface, \ie, for a one-dimensional metal grating. 

More generally, we can introduce a spatial cutoff frequency $\kappa>0$ such that $c_k=0$ for all $k<\kappa$. In this case, the refractive index will approach the vacuum limit $n=1$ according to $e^{-\kappa v}$ for $v\to \infty$. Hence, the metal surface $v(u)$ can be almost perfectly cloaked using a finite-size refractive-index profile with an extent comparable to only $2\pi/\kappa$. 

\begin{figure}[h]
	\begin{center}
		\rule{0.5\textwidth}{0mm}\includegraphics[width=\legwidth]{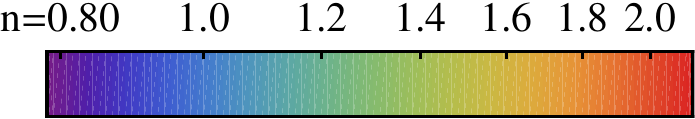}\\
		\includegraphics[width=0.49\textwidth]{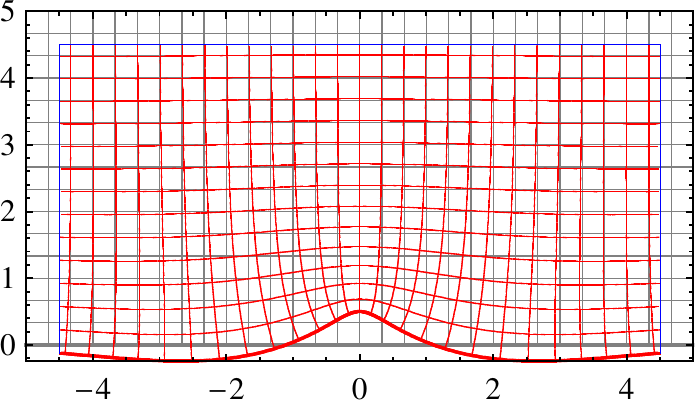}
		\includegraphics[width=0.49\textwidth]{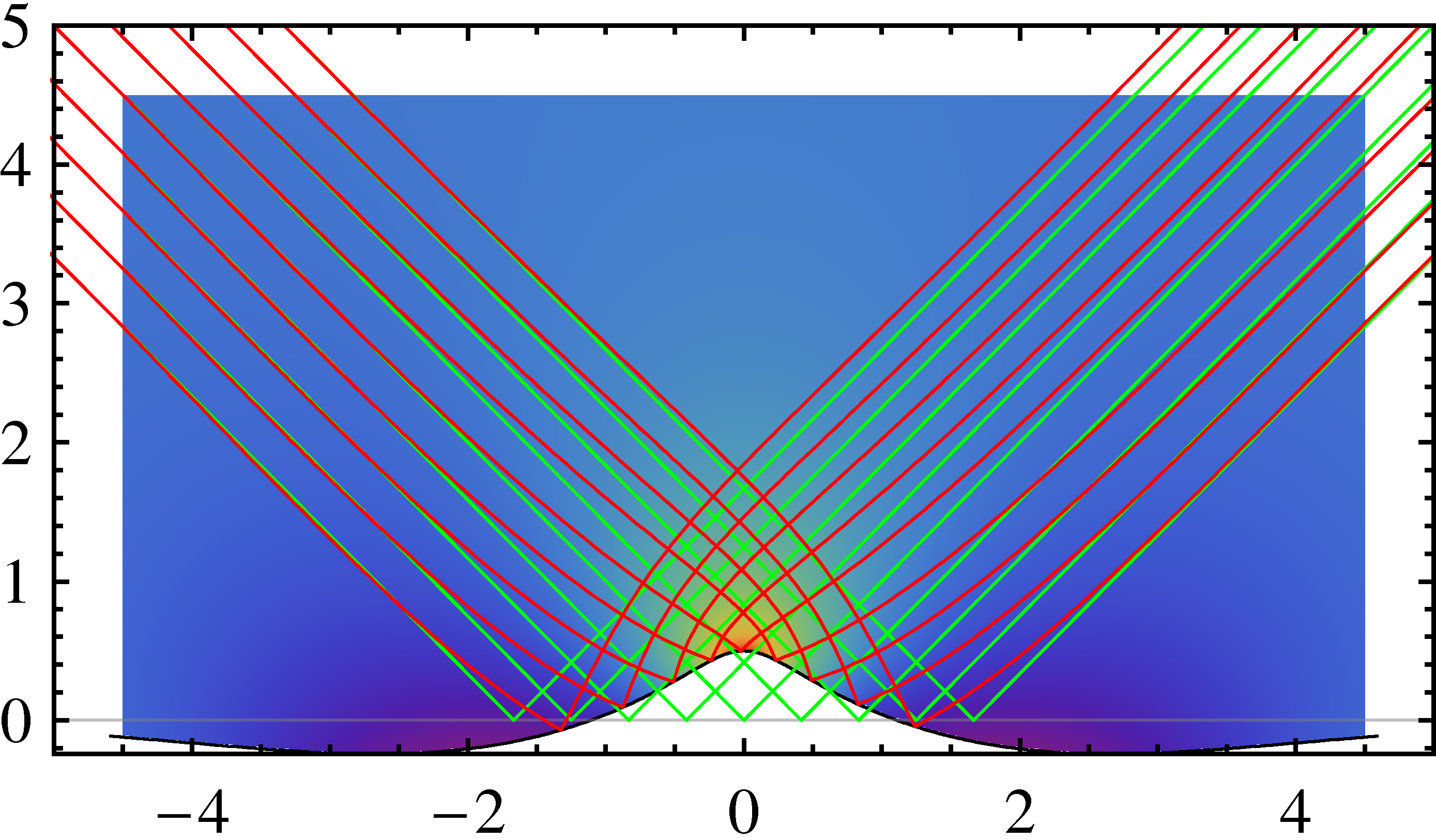}
	\end{center}
	\caption{Left: illustration of the original coordinates $(x,y)$ (gray) and the transformed coordinates $(u,v)$ (red) resulting from Eq.~\eqref{eq:GaussianBumpCM2} with $w=2$, $h=0.5$, and $\kappa=0.5$. All coordinates are in normalized units. Right: refractive-index profile and selected rays. The green rays correspond to vacuum and the mirror plane indicated in gray, the red ones to the finite cloaking structure and the ground-plane shape shown in black. The cloak has a size of $9\times4.5$ normalized units. Outside of it, we assume vacuum (white). The color scale is logarithmic, ranging from 0.78 to 2.14.}
	\label{fig:GaussianBumpExp}
\end{figure}

As an example we apply this cutoff to the Gaussian map resulting from Eq.~\eqref{eq:GaussianCoeff}, with which we had started our discussion. Interpreting the cutoff as a shift on the spatial modes, this leads to
\begin{equation}
	\label{eq:GaussianBumpCM2}
	f(z) = z+\int_{0}^{\infty} c_k e^{\ii (k+\kappa) z}\text{d}k
	= z + \ii h e^{-(z/w)^2} \left[ 1+\erf(\ii z/w) \right] e^{i \kappa z}\,.
\end{equation}
Far from the bump, the refractive-index profile derived from this transformation decays according to
\begin{equation}
	n(u+\ii v=\rho e^{\ii\varphi}) =
	1 + \frac{h w \kappa \sin(\varphi-\kappa\rho\cos\varphi)}{\sqrt{\pi}}
	\times \frac{e^{-\kappa\rho\sin\varphi}}{\rho}
	+ \mathcal{O}\left(\frac{e^{-\kappa\rho\sin\varphi}}{\rho^2}\right).
\end{equation}
The behavior of this modified Gaussian bump is illustrated in Fig.~\ref{fig:GaussianBumpExp}. In particular we note in the left panel that the transformed coordinates $(u,v)$ converge rapidly towards the Cartesian ones $(x,y)$ when moving away from the mirror plane. In the right panel this finding translates into much smaller beam displacements than those discussed in Ref.~\cite{Zhang2010} (and in our Fig.~\ref{fig:GaussBump}) between the green rays (vacuum, reflected off of a planar mirror) and the red rays (reflected off of the curved mirror shown in black) when using such a finite-size cloak. With a further increase in cloak size these shifts disappear \emph{exponentially}, as opposed to the polynomial decrease observed for $\kappa=0$. We have found the same behavior for rays impinging under different angles (not depicted). 

\begin{figure}[h]
	\begin{center}
		\rule{0.5\textwidth}{0mm}\includegraphics[width=\legwidth]{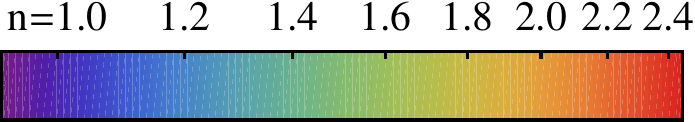}\\
		\includegraphics[width=0.49\textwidth]{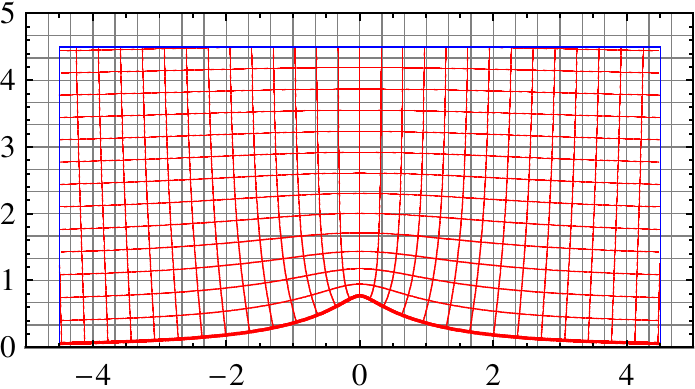}
		\includegraphics[width=0.49\textwidth]{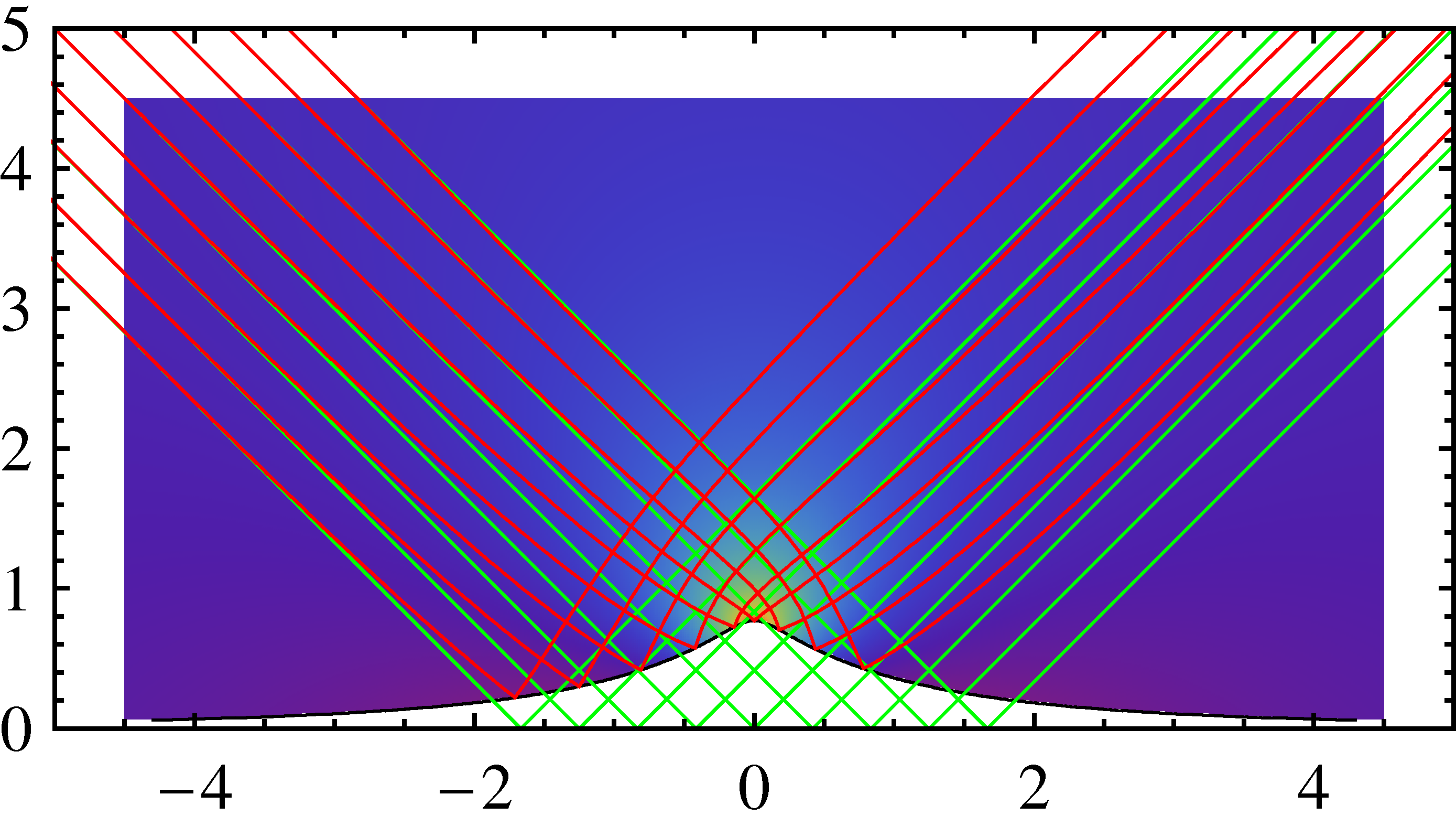}
	\end{center}
	\caption{Left: illustration of the original coordinates $(x,y)$ (gray) and the transformed coordinates $(u,v)$ (red) resulting from the cloak of Ref.~\cite{Zhang2010b}, {\it i.e.}, for the conformal map $z \mapsto f(z)=z-1/(z+\ii \delta\!y)$ with $\delta\!y=1.3$. All coordinates are in normalized units. Right: refractive-index profile and selected rays. The green rays correspond to vacuum and the  mirror plane at $y=0$, the red ones to the finite cloaking structure and the ground-plane shape shown in black. The cloak has a size of $9\times4.5$ normalized units. Outside of it, we assume vacuum (white). The color scale is logarithmic, ranging from 0.92 to 2.45.}
	\label{fig:ZhangBump}
\end{figure}

It is instructive to compare our results with those of Ref.\,\cite{Zhang2010b}, which has used the conformal map $z \mapsto f(z)=z-1/(z+\ii \delta\!y)$ and a mirror plane at fixed height $\delta\!y=1.3$; in our notation this cloak results from $c_k=\ii \exp(-k \delta\!y)$. Corresponding results for a finite-size cloaking structure with a height of 4.5 normalized units are shown in Fig.~\ref{fig:ZhangBump} -- allowing for direct comparison with Figs.~\ref{fig:GaussBump} and~\ref{fig:GaussianBumpExp}. The lateral beam displacement highlighted in Ref.~\cite{Zhang2010} is very pronounced for this cloak, which, as in Fig.~\ref{fig:GaussBump}, is due to the presence of strong components $c_k$ at small spatial frequencies $k$ and thus a slowly decaying refractive-index profile:
\begin{equation}
	n(u+\ii v=\rho e^{i\varphi}) = 1-\frac{\cos(2\varphi)}{\rho^2}+\mathcal{O}(\rho^{-3}).
\end{equation}

Finally, we note that refractive-index profiles for cloaking further ground-plane shapes $v(u)$ can easily be obtained with our analytical approach. Textbook examples of conformal maps representing circular, rectangular, and triangular bumps, however, lead to infinities in the required refractive-index profiles. In contrast, as long as the shapes $v(u)$ are smooth and do not exhibit any kinks, the resulting conformal maps and refractive-index profiles are smooth as well and represent realistic proposals for experimental cloaks.

\section{Conclusion}

To summarize, we have introduced a class of conformal versions of the quasi-conformal carpet cloak previously introduced by Li and Pendry. We have obtained exact analytical mathematical forms for the refractive-index profiles of usual bumps as well as of other shapes, \eg, of gratings. Our analytical formulas can simply replace nontrivial numerical calculations along the lines of the quasi-conformal mapping. This step considerably eases working with these refractive-index profiles in practice. The analytical forms also allow us to study the asymptotic behavior of the refractive-index profiles (\eg, polynomial or exponential). This aspect is important for assessing and optimizing the performance of 
{\it finite-size} cloaks as blueprints for experiments. In this regard, we obtain much smaller lateral beam displacements for certain metal profiles than previous quasi-conformal \cite{Zhang2010} as well as previous conformal maps \cite{Zhang2010b}.

\section*{Acknowledgements}

We thank Nicolas Stenger and Tolga Ergin (Karlsruhe) for discussions and for a critical reading of the manuscript. M.\,W. acknowledges support by the Deutsche Forschungsgemeinschaft (DFG) and the State of Baden-W\"urttemberg through the DFG Center for Functional Nanostructures (CFN) within subproject A\,1.5. The project PHOME acknowledges the financial support of the Future and Emerging Technologies (FET) programme within the Seventh Framework Programme for Research of the European Commission, under FET-Open grant number 213390. The project METAMAT is supported by the Bundesministerium f\"ur Bildung und Forschung (BMBF).

\end{document}